# A single-electron transistor made from a cadmium selenide nanocrystal

**David L. Klein[a,c], Richard Roth[b,c], Andrew K.L. Lim[b,c], A. Paul Alivisatos[b,c], and Paul L. McEuen[a,c]**

[a]Department of Physics, University of California,
[b]Department of Chemistry, University of California, and
[c]Molecular Design Institute, Lawrence Berkeley National Laboratory, Berkeley California 94720.

**Using colloidal chemistry techniques, it is possible to routinely create semiconductor nanocrystals[1,2] whose dimensions are much smaller than those that can be realized using lithographic techniques.[3-6] Their sizes can be systematically varied to study quantum confinement effects or to make novel electronic/optical materials with tailored properties.[7-9] Preliminary studies of both the electrical[10-13] and optical properties[14-16] of individual nanocrystals have been recently performed. These studies clearly show that a single excess charge on a nanocrystal can dramatically influence its properties. Here, we present measurements where the charge state of the nanocrystal can be directly tuned. We describe electrical transport in a single electron transistor made from a colloidal nanocrystal. A voltage applied to a gate changes the number of charge carriers on the nanocrystal and allows the measurement of the energy for adding successive charge carriers. As studies of lithographically patterned quantum dots[3-6] and small metallic grains[17] have shown, such measurements are invaluable in understanding the energy level spectra of small electronic systems.**

Figure 1a shows an idealized schematic of the device. 5.5 nm diameter CdSe nanocrystals[18,19] are bound to two closely spaced Au leads using bifunctional linker molecules[20,21]. The leads are fabricated on a degenerately doped silicon wafer which is then used as a gate to tune the charge state of the nanocrystal under study. Figure 1b shows a field emission scanning electron micrograph of a completed device. A number of nanocrystals appear to be in the ~ 5 nm gap between the electrodes.

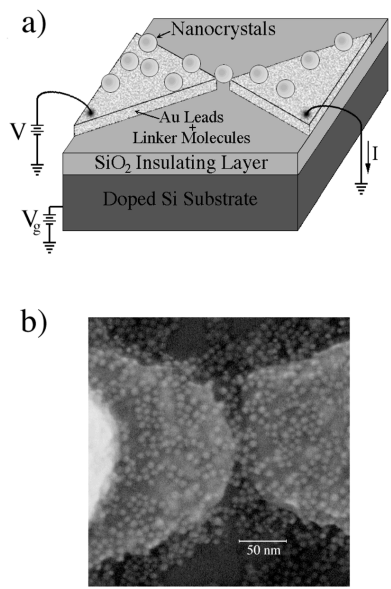

Figure 1. (a) Schematic of the device. Nanocrystals sit atop two closely spaced leads. Transport measurements probe the nanocrystal that best bridges the gap between the leads. The electrochemical potential of this nanocrystal can be tuned by applying a gate voltage to the substrate. (b) Field emission scanning electron micrograph of a set of 13 nm thick leads separated by a ~ 5 nm gap onto which 5.5 nm diameter CdSe nanocrystals have been deposited. A combination of optical and electron beam lithography coupled with shadow evaporation is used to define these leads. The leads are patterned on a degenerately doped wafer on which a 53 nm thick oxide barrier has been grown. The nanocrystals have a mean diameter of 5.5 nm and a 5% coefficient of variation and are attached to the leads with 1,6-hexanedithiol. One end of these linear dithiol molecules binds to Au leads and the other binds to the nanocrystals to form a 1.2 nm thick insulating barrier. Further fabrication details are described in Ref. 11.

Most devices, including the particular junction shown in Figure 1b have immeasurably high impedance (R > 100 GΩ). Only about one in twenty have a measurable conductance, typically in the 10 MΩ – 1 GΩ range. Devices of this type typically behave as though transport is occurring through a single nanocrystal[11], as we demonstrate below. This may initially seem surprising, since Figure 1b indicates that the number of nanocrystals in the junction region is quite large. However, tunneling through the linker molecules has an exponential decay length of less than 1 Å[11,20,22]. As a result, only a well-placed nanocrystal (within 2 nm of each lead) can contribute to conduction.

Figure 2 shows the linear response conductance G of a device measured at T = 4.2 K as a function of the gate voltage, $V_g$, applied to the substrate. As the gate bias is raised, the conductance grows to a peak and then declines back to zero. In the insets to Figure 2 the current I flowing through the nanocrystal is plotted as a function of the voltage V applied between the two leads at two fixed values of $V_g$. The I-V taken at a $V_g$ away from the linear response peak shows a suppressed conductance at small V while the I-V characteristic taken at the center of the peak has a finite linear conductance for small V. Figure 3 shows a map of the differential conductance of the device, plotted as a gray scale, as a function of both V and $V_g$.

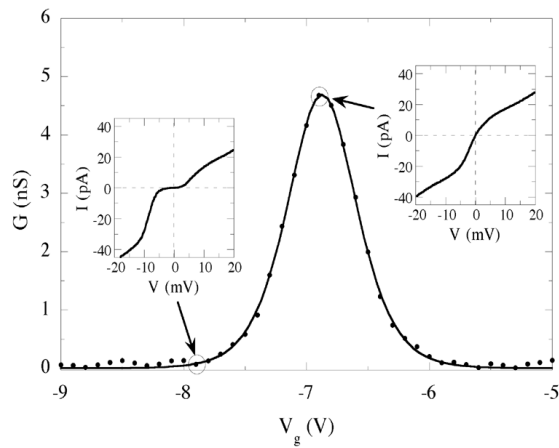

Figure 2. Conductance G versus gate voltage $V_g$ for a single nanocrystal transistor measured at T = 4.2 K. The conductance shows a peak when the charge state of the nanocrystal changes by one electron. The dots are the measured values; the solid curve is a fit to the data using the standard Coulomb blockade model[6] with a temperature T = 5 K. Insets: I-V characteristics measured at the gate voltages indicated.

These measurements can be understood using ideas borrowed from studies of lithographically patterned dots[3,5,6]. The conductance peak in Figure 2 is a Coulomb oscillation. The suppressed conductance on either side of the peak is a consequence of the finite energy required to add (remove) an electron to (from) the nanocrystal in its ground state. This energy is analogous to the electron affinity (ionization energy) of a molecule. The peak occurs when the two charge states of the nanocrystal have the same total (including electrostatic) energy and an extra electron can therefore hop on and off the nanocrystal at no energy cost. The maximum size of the gap in the I-V curves off of the peak provides a measure of the addition (removal) energy for electrons, as we discuss further below.

Figure 4a is a schematic energy level diagram of the nanocrystal with N electrons at a gate voltage midway between two Coulomb oscillations. The electrochemical potential for adding the $(N+1)^{th}$ electron to the nanocrystal is denoted $\mu_{N+1}$. The energy difference $\Delta_{N+1}$ between $\mu_{N+1}$ and $\mu_N$ is referred to as the addition energy In the simplest model, referred to as Coulomb blockade model[6], the addition energy is given by:

$$\Delta_{N+1} = \mu_{N+1} - \mu_N = U + \Delta E,$$

where U is the Coulomb interaction energy between any two electrons on the nanocrystal and $\Delta E = E_{N+1} - E_N$ is the energy level spacing to the next unoccupied single-particle eigenstate.

When the linear response conductance is zero, a finite V can adjust the Fermi energies of the leads relative to the nanocrystal to add or remove an electron from the nanocrystal. The Fermi level of the left lead, equal to -eV, can either rise above $\mu_{N+1}$ or fall below $\mu_N$, allowing current to flow. The range of V for which the conduction is zero is referred to as the Coulomb gap. This gap varies as a function of $V_g$. The maximum size of the Coulomb gap for N electrons on the nanocrystal, i.e. the maximum voltage V that can be applied without current flow, is a direct measurement of the addition energy $\Delta_{N+1}$ for the next electron. This is illustrated in Figure 4b.

The evolution of the Coulomb gap with $V_g$ can be seen in figures 3a and 3b, where the differential conductance dI/dV is plotted as a function of both $V_g$ and V. The Coulomb gap produces the light colored diamond-shaped regions in these figures. The Coulomb gap is zero at a Coulomb oscillation, it grows to a maximum size approximately halfway between two oscillations, and then decreases again to zero at the next Coulomb oscillation. Figure 3c shows a schematic evolution of the Coulomb gap, along with the inferred addition energies $\Delta_N$ for successive electrons. These energies range from approximately 15 to 30 meV.

Similar although less complete results have also been obtained in two other gated samples. In addition, I-V measurements on eight other samples without a gate have been performed, some of which have been presented previously[11]. The addition energies determined for these samples are typically between 30 and 60 meV.

In the Coulomb blockade model, the addition energy for adding successive carriers is $U+\Delta E$. A straightforward electrostatic estimate for U, including the screening effects of the metallic leads, yields U ~ 50 meV for a 5.5 nm diameter metallic sphere[11]. The level spacing $\Delta E$ depends on the type of charge carriers and is predicted to be ~ 100 meV for electrons and ~ 10 meV for holes[23]. The measured values of 15-60 meV for the addition energies are thus consistent with, but somewhat smaller than, expectations (~ 50-60 meV) for individual holes on a CdSe nanocrystal.

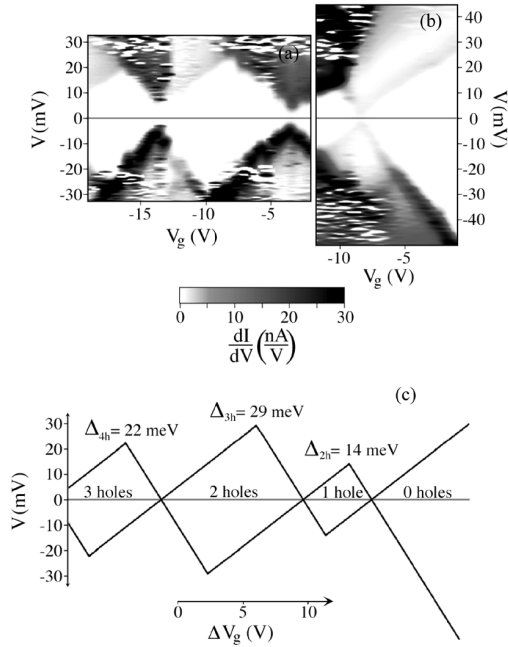

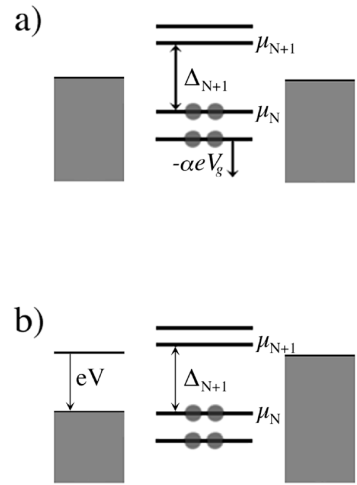

Figure 3. (a),(b) Composite gray scale plots of the differential conductance dI/dV of a CdSe nanocrystal plotted as a function of both $V_g$ and V. The white diamond-shaped regions correspond to the Culomb gap. During the course of taking this data, several "switches" occurred where the entire characteristic shifted along the $V_g$ axis. These switches are ubiquitous in Coulomb blockade measurements and are thought to result from sudden changes in the local electrostatic environment. One such change occurred between the acquisition of figures 3a and 3b. During the acquisition of figure 3a, two smaller shifts occurred. To account for this, some parts of this data have been translated along the $V_g$ axis. (c) A schematic illustration of the data from (a) and (b) indicating the inferred number of holes on the nanocrystal as a function of $V_g$ and the addition energy for successive holes.

Figure 4: (a) Energy level diagram for a nanocrystal with N electrons at a gate voltage midway between two Coulomb oscillations. The electrochemical potential of the N electron nanocrystal is denoted $\mu_N$, while the electrochemical potential of the N+1 electron nanocrystal is denoted $\mu_{N+1}$. (b) The application of a finite bias V can overcome the Coulomb gap. The maximal V applied before conduction occurs is equal to the addition energy $\Delta_{N+1}$ for the (N+1)$^{th}$ electron.

We now address an intriguing aspect of the data. Note that on the right side of figure 3 the Coulomb gap continues to grow with increasingly positive $V_g$. For even larger gate voltages than shown in the figure (up to 40 V) the Coulomb gap exceeds 150 mV and there is no evidence for any more Coulomb oscillations.

This behavior can be understood if electrons are being added to the valence band of the nanocrystal with increasingly positive $V_g$. In this case, for some sufficiently positive $V_g$, every state in the valence band is filled, and the next extended state of the nanocrystal lies in the conduction band, 2 *eV* higher in energy. In this interpretation, the large gap at positive $V_g$ corresponds to a completely filled valence band. Starting on the right side of Figure 3 and decreasing $V_g$, the first Coulomb oscillation of Figure 3 then corresponds to the removal of the first electron from, or, alternatively, the addition of the first hole to, the valence band. A further decrease in $V_g$ adds additional holes. The addition energies for the 2$^{nd}$, 3$^{rd}$, and 4$^{th}$ holes are: $\Delta_2 = 14 \pm 2$ meV, $\Delta_3 = 29 \pm 3$ meV, and $\Delta_4 = 22 \pm 2$ meV, as is indicated in Figure 3(c).

Note that in the Coulomb blockade model, the energy $\Delta_2$ for adding the second hole to the nanocrystal is simply the Coulomb interaction U between holes, since the lowest-lying hole state is doubly degenerate. The energy $\Delta_3$ for adding the third hole, on the other hand, is U+$\Delta$E, where $\Delta$E is the difference between the ground and first excited single particle state. The addition energy for the fourth hole $\Delta_4$ is again U since the first excited state is also doubly degenerate. The results given above are somewhat consistent with this scheme — e.g. $\Delta_3$, predicted to be U +$\Delta$E, is greater than $\Delta_2$, which is predicted to be U.

Discrepancies remain, however. The energy for adding the second hole, $\Delta_2 = 14$ meV is significantly smaller than the values of U ~ 50 mV obtained from simple electrostatic estimates or from measurements of other samples. In addition, the measured $\Delta_4$ is greater than $\Delta_2$, while in the Coulomb blockade model they should be the same. The origin of these differences remains unclear, although the effects of exchange and correlations are expected to be important for these few-hole systems. We also emphasize that the large gap at positive $V_g$ has only been seen in one device. To confirm that this gap is associated a filled valence band and to verify the inferred hole addition energies, it is necessary to repeat these measurements on other samples.

This work represents a new type of spectroscopy for single nanocrystals. Unlike optical measurements, where electron-hole pairs are created, these measurements probe the energy for adding a single type of charge carrier. We hope that these results will stimulate new calculations of the hole addition energies for a CdSe nanocrystal that include the effects of exchange, correlations, and screening by the metallic electrodes. Future measurements will investigate how the ground state and excited state properties vary with the size, shape, and composition of nanocrystals as well as the composition of the leads.

**Acknowledgments.** This work was supported by the U.S. Department of Energy and by the Office of Naval Research.

Correspondence should be addressed to P.L.M.
(email mceuen@physics.berkekely.edu).